# Statistics for Building Synthetic Power System Cyber Models

Morayo Soetan
Dept. of Electrical & Computer Engineering
Texas A&M University
College Station, Texas
MorayoS1@tamu.edu

Zeyo Mao
Dept. of Electrical & Computer Engineering
Texas A&M University
College Station, Texas
zeyumao2@tamu.edu

Katherine Davis
Dept. of Electrical & Computer Engineering
Texas A&M University
College Station, TX, USA
katedavis@tamu.edu

*Abstract*—A realistic communication system model is critical in power system studies emphasizing the cyber and physical intercoupling. In this paper, we provide characteristics that could be used in modeling the underlying cyber network for power grid models. A real utility communication network and a simplified inter-substation connectivity model are studied, and their statistics could be used to fulfill the requirements for different modeling resolutions.

## I. Introduction

Modern power grids have become increasingly interdependent on their communication networks. Distributed energy resource (DER), virtual power plant (VPP), remedial action scheme (RAS), etc., all these emerging technologies requires the existence of a robust underline cyber system where the communication network plays an important role. Research in these fields and many other power systems (e.g., cyber-security, distributed control) normally either study the pattern of the interaction between the power (physical) system and the communication (cyber) system, or give an assumption of this interaction, both require an appropriate model of the underlying communication network. Unfortunately, most real utility communication network models belong to CEII (Critical Electric Infrastructure Information), thus they are usually confidential and hard to access. In [1] the authors propose a cyber network model with snowflake or radial topology for the synthetic Texas power grid, where the utility control center is directly connected to each managed substation, and there are no inter-substation communication links. In [2] the authors use the same topology as the studied power grid model to represent the communication network topology, which is largely due to the fact that the studied real power system uses the optical ground wire (OPGW) as its major communication media between substations. These models are useful in their applications; however, their modeling method might not work as a general-purpose approach due to some limitations caused by their assumption. For example, routing optimization could not be done in a communication network with radial topology. Also, for most cases, the communication network is unlikely to be - the same as the power network, because OPGW is normally used in the extra-high voltage (EHV) system, and the prevalence of other communication media in the power system and do not go along with the transmission lines.

To the best of the author's knowledge, few pieces of literature extract characteristics from a real utility communication system and creating a realistic synthetic cyber network using these statistics. This should be changed since the dependence on the communication system is growing as the grid is becoming "smarter". Recently [3] provides us a valuable topology model from a realistic utility communication system, as well as analyzing some structural metrics, such as degree and centrality, which are quite useful in generating synthetic networks. The model they studied has different types for node (station) and degree (communication link), thus it might be overkill for studies where a simplified but realistic communication connectivity model is preferred. For instance, this communication connectivity model could be used in studies where the availability of the generation resources is a major concern, or studying the impact of some threat models [4], or where the cyber mitigation is involved, like the routing optimization for specific scenarios. As a comparison, the detailed model could be used in situations where the communication links are not negligible due to their unique characteristics and abilities to handle special events. In this paper, we want to extract enough useful statistics from this original topology model and a simplified model, which later will be used to design synthetic communication networks for different resolution levels. The contribution of this paper mainly consists of 1) providing an efficient way to extract the communication connectivity from the original detailed model 2) analyzing both detailed and simplified models to provide statistics to support generating synthetic cyber networks for power grid models. The paper will be organized as follows: section II will introduce both models and discuss the simplifying process; section III will define the studied metrics and present the results; the remaining sections will conclude the paper and discuss the followed future work.

## II. Network Models

### A. Detailed Network Model

The detailed inter-substation network model was manually parsed from [3] and then re-construct using the network analysis tool NetworkX [5] to build the standard graph model. The communication network consists of 333 nodes represented by microwave stations, transmission stations, generation stations, offices, very-high frequency repeaters (VHF), ultra-high frequency repeaters (UHF), control centers, and other grid components. The network has 369 links represented by microwave, fiber, power-line communication (PLC), and leased links with 15 parallel cases.

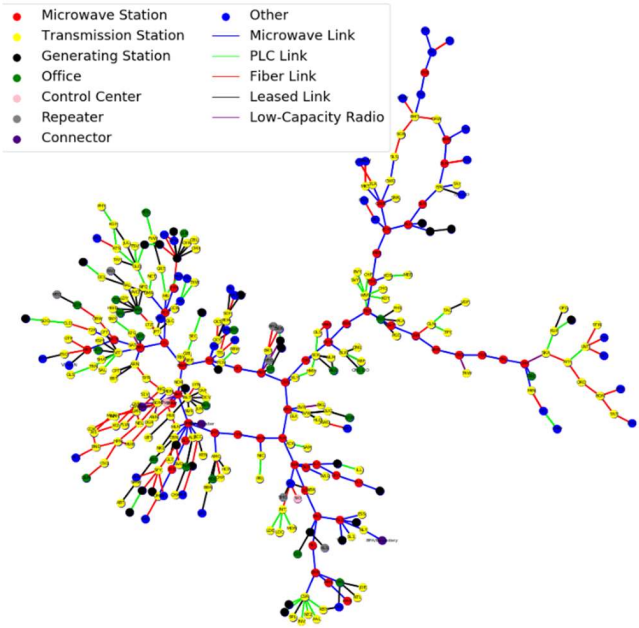

Fig 1. The Detailed Inter-Substation Network Topology of a Realistic Utility Communication System

In the original communication network, 11 island nodes, nodes disconnected from the main structure, were present. Assuming these nodes were either defuncted stations or stations in development, none of the island nodes were included in the reconstruction of the detailed network. As a result, the studied network was fully-connected. When estimating the detailed network's link lengths, the corresponding transmission system [6] did not mirror the communication network. As a result, the provided link lengths were negligible in the detailed network and the links seen in Figure 1 solely represent the connectivity between stations.

### B. Simplified Connectivity Model

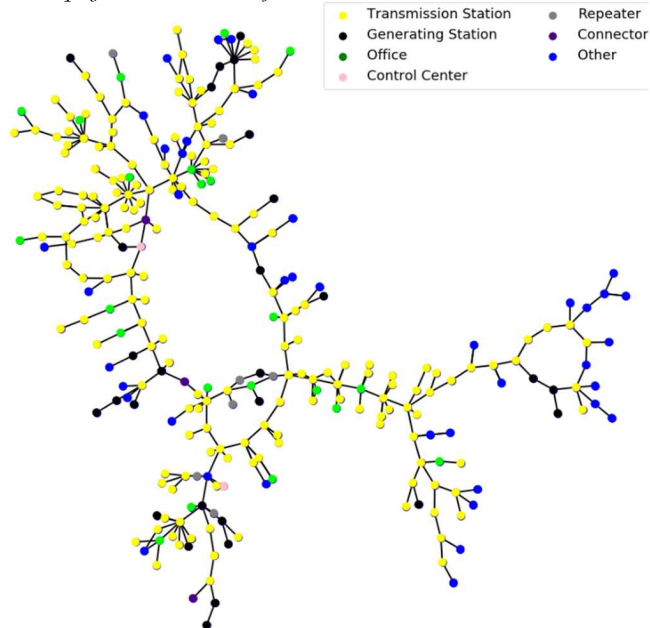

Fig 2. Simplified Inter-Substation Connectivity Model

Depicted in Figure 2, the simplified model represents the transmission-level inter-substation connectivity, which is extremely useful in some power system control and security problems where a high-level reachability matters. This model could be used in applications where the type of communication link is neglectable.

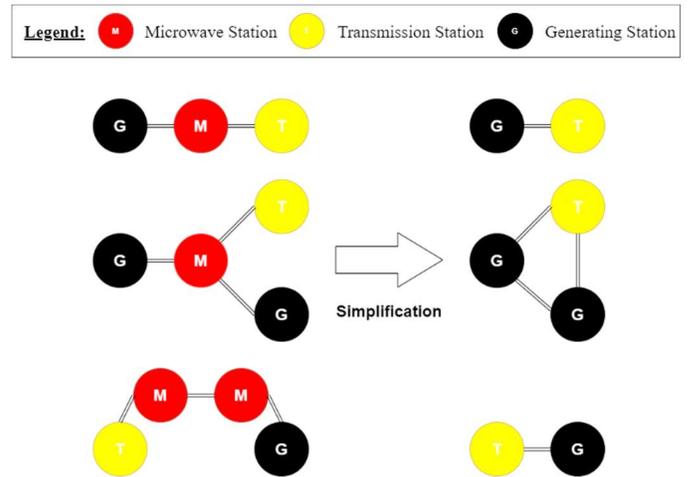

Fig 3. Visual representation of simplification process used to produce the Simplified Inter-Substation Network.

As shown in Figure 3, to create the simplified version of the detailed communication network, each microwave station's neighbor nodes were circular linked and the microwave station was removed. If the microwave station possessed one degree, the microwave station was removed without performing circular linking. This process was repeated until all of the microwave stations were removed from the network. Since the simplification process required us to remove and create degrees between nodes, the degree types demonstrated in Figure 1 were not included in this model. As a result of the simplification process, the simplified communication network had 279 nodes and 333 degrees. Alike to the detailed network, the simplified network was fully-connected because the network was extracted from the fully-connected detailed network and the simplification process did not produce island nodes.

### III. STATISTICS FROM CYBER MODELS

As mentioned in the introduction, the objective of the paper is to extract useful statistics from a detailed and simplified communication network devised from a real utility communication network. In this section, we will define and analyze the usability of degree type distributions, average degree load, primary shortest pathlength distributions, and edge betweenness centrality.

### A. Degree Type Distributions

By providing information a medium to transmit between nodes, degrees are the foundation for all communication networks. Statistics collecting quantifiable data from a network's degrees, including degree type distribution, have the potential to accurately assess the characteristics of a communication network. Thus, they are also important validation metrics for designing synthetic communication networks. There is a variety of communication links used to deliver the packet flow. Each variation has unique characteristics influenced by their respective equipment, materials, and communication mediums. To understand the significance of degree type in communication networks, a degree type distribution, representing the relationship between the degree and node types, was created for the detailed network.

TABLE I. Detailed Inter-Substation Network's Degree Type Distribution

| Node Type | Degree Type | | | | |
|---|---|---|---|---|---|
| | Microwave | PLC | Fiber | Leased | Radio |
| Microwave | 22.8% | 0.1% | 1.4% | 0.5% | 1.1% |
| Transmission | 6.8% | 11.9% | 13.7% | 15.9% | 2.0% |
| Generating | 1.4% | 2.0% | 1.4% | 1.8% | 0.1% |
| Office | 0.4% | 0.1% | 0.4% | 4.9% | 0.0% |
| Control Center | 0.1% | 0.0% | 0.4% | 0.0% | 0.0% |
| Repeater | 0.0% | 0.1% | 0.3% | 0.5% | 0.3% |
| Connector | 0.3% | 0.0% | 0.5% | 0.0% | 0.0% |
| Other | 1.4% | 1.1% | 4.7% | 1.6% | 0.0% |
| Total Percentage | 33.1% | 15.4% | 22.8% | 25.2% | 3.5% |

Briefly mentioned in the introduction, some communication degrees such as microwave, low-capacity radio, and leased degrees, tend to not accompany transmission lines. Conversely, PLC and fiber degrees use transmission lines existing infrastructure to deliver information. Considering the tendencies of the degree types, the ratio of PLC and fiber degrees to the total number of degrees, known as the PLC-Fiber ratio, was calculated to quantify the influence of transmission lines in the communication network's design. The measured PLC-Fiber ratio of the detailed network was 38.21%, suggesting the detailed network's design made use of several sections of the transmission system's infrastructure. This suggestion was supported by the high percentage of transmission stations connected to PLC and fiber degrees shown in Table 1. Moreover, the radial topology branching from the detail network's main stem possessed a significant share of transmission stations, PLC degrees, and fiber degrees, which are indicators of transmission lines. With the PLC-Fiber ratio, a synthetic communication network's transmission system dependence could be accurately measured and assessed.

B. *Average Degree Load*

Derived from a network's degree count distribution, a node type's average degree load (ADL) statistic averages the degree count for a specified node type. Since microwave, transmission, and generating stations constitute a majority of the networks' nodes, their respective ADLs were analyzed to gain insight into the network's design.

TABLE II. Detailed and Simplified Inter-Substation Networks' Average Degree Loads

| Node Type | Average Degree Loads | |
|---|---|---|
| | Detailed | Simplified |
| Microwave Station | 3.537 | -- |
| Transmission Stations | 2.108 | 2.54 |
| Generating Stations | 1.75 | 2.286 |
| Office | 1.955 | 2.0455 |
| Control Center | 2 | 2 |
| Repeater | 1.289 | 1.428 |
| Connector | 2 | 2.333 |
| Other | 1.585 | 1.683 |

For the detailed network, the microwave station possessed an ADL value of 3.5. The value implies that many microwave stations were points of divergence for nodes/node clusters. Moreover, the detailed network's transmission and generating stations' relatively low ADLs, comparative to the microwave's ADL, implies that many of these nodes were on the branching and ends of node clusters. Depicted in Figure 1, the microwave stations formed the detailed network's main stem, while the transmission and generating stations resided heavily on the branching of the network, supporting the aforementioned implications. Validated by Figure 2, the simplified network's transmission and generating station's ADLs indicate that the majority of transmission and generating stations were points of divergence or resided in node clusters. Demonstrated from the analysis, measuring a network's node types' ADLs could prove useful in network modeling by revealing the neighboring tendencies of each node type.

C. *Routing Path*

Observable in Figures 1 and 2, the control centers accessed substations using the network's various paths. Thus, when designing a communication network one critical procedure is to plan/optimize routing paths. The most widely used routing approach is shortest path routing (SPR), which determines the minimal path and pathlength between the nodes and control centers. Since a communication network's efficiency is heavily reliant on pathlength, the statistics and distributions produced from SPR could accurately measure network efficiency, thus they are also important validation metrics for synthesizing cyber networks.

When performing SPR on the networks, Dijkstra's algorithm was utilized and the degrees were assigned a uniform weight. [7] describes Dijkstra's algorithm as a process used to record the known shortest pathlength between each node to the source node and updates the pathlengths when shorter paths are discovered. If the shortest path between a node and the source node is achieved, the node is added to the path and marked as "visited". The process is repeated until all nodes are visited. For each node, two shortest paths from the node to control centers 1 & 2 were built using Dijkstra's algorithm. To determine each node's primary shortest pathlength (PSL), the node's shortest paths' pathlengths were compared to find the smallest pathlength value. Then, the smallest pathlength was assigned as the node's PSL. If both paths' pathlengths were equivalent the given node's PSL was set to the value of equivalence. This process was replicated for each node in the respective networks.

---

**Algorithm 1** Pseudocode for Primary Routing Path

1: **function** *Primary_Path(Node_list, Control_list)*
    ‣ Node list is a list of each nodes' name
    ‣ Control list is a list of Control Centers' names
2:     *primary_length* = empty one-dimension list with *Node_list's* length
3:     **for** node *nod* in *Node_list* **do**
4:         Create instance of current shortest pathlength *short*
5:         **for** control *cont* in *Control_list* **do**
6:             Calculate *path_nodes* and *path_length* between node and cont using Dijkstra's algorithm
7:             **if** *cont* is first index in *Control_list* **then**
8:                 set *short* to *path_length*
9:             **elif** *short* is greater than path_length **then**
10:                set *short* to *path_length*
11:        Populate *primary_length* with *short* at *nod*'s index in *Node_list*
12:    **return** *primary_length*
13: **end**

---

Fig 4. Pseudocode calculating each node's primary routing path

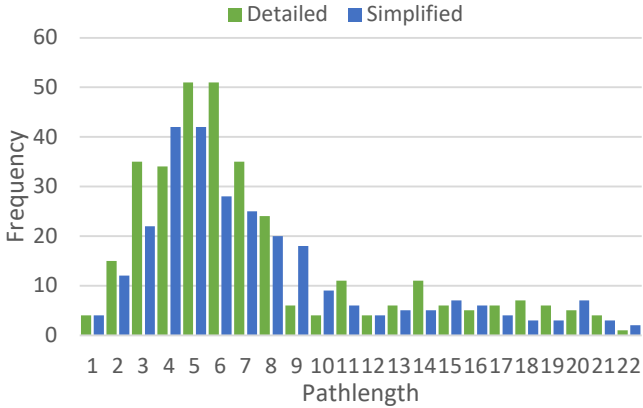

Fig 5. Primary Shortest Pathlength Distribution of Detailed and Simplified Inter-Substation Networks

Using the gathered PSL data, a primary shortest pathlength distribution was produced for each network. To gain an understanding of the distributions, a value known as Pearson's first coefficient of skewness was calculated. In [8], an equation for Pearson's first coefficient makes use of a population's mean, mode, and standard deviation to measure skewness.

$$skewness\ coefficient = \frac{\bar{x} - mode}{\sigma} \quad (1)$$

A Pearson's first coefficient describes three types of distributions: A positive coefficient represents a positively-skewed distribution, a negative coefficient represents a negatively-skewed distribution, and a zero coefficient represents a normal distribution. It was determined that the detailed and simplified networks' primary shortest pathlength distributions' Pearson's first coefficient of skewness were 0.419 and 0.642. Indicated by the coefficients and Figure 5, the primary shortest pathlength distributions were skewed positively, revealing the networks' prioritization of smaller pathlengths. The distributions' significant share of small pathlengths heavily influenced the distributions' mode and Average Path Length (APL), statistics known to measure network efficiency. The detailed and simplified networks' skewness coefficient suggests that optimizing a network's primary shortest pathlength distribution's skewness in the positive direction could effectively increase overall network efficiency.

*D. Edge Betweenness Centrality*

Synonymous with communication degrees, edges are the main factor in determining the overall structure of networks. Due to edges' strong influence on the network, statistics that observe the behavior of edges, such as Edge Betweenness Centrality (EBC), could serve as network efficiency indicators and edge type assignment guidelines. EBC is a metric used to quantify a specified degree's importance in a network. To measure a degree's EBC, SPR is utilized to measure the shortest path(s) between two nodes. With a selected degree, the degree's total appearance in the shortest paths(s) is divided by the total number of observed shortest paths. The aforementioned process is replicated for all possible combinations of two nodes. Then, a summation is taken, resulting in the selected degree's EBC. The NetworkX analysis tool's edge betweenness centrality built-in function proved very useful in finding edge's EBC, reducing time and work needed for computation. To determine the usefulness of EBC in synthetic cyber networks, the detailed network's degrees' EBC were calculated and averaged with their respective degree types. The resulting values, known as Average Edge Betweenness Centrality (AEBC), revealed each degree types' structural importance in the communication network.

TABLE III. Detailed and Simplified Inter-Substation Networks' Average Edge Betweenness Centrality

| Degree Type | Average Edge Betweenness Centrality |
|---|---|
| Microwave | 4155.23 |
| PLC | 468.9 |
| Fiber | 668.85 |
| Leased | 391.24 |
| Low-Capacity Radio | 1320.5025 |

Evident in Table 3, the microwave degrees possessed the largest AEBC by a sizable margin. The vast difference between the microwave station's AEBC to the other edge types' AEBC suggested that the network structural integrity was very dependent on microwave edges. Unexpectedly, the low-capacity radio possessed a relatively large AEBC, despite it being the least used edge type in the network. Since microwave and low-capacity radios are forms of wireless transmission, it was inferred that the network primarily relied on wireless communication to transmit data due to the two link types' high AEBC values. Using this inference as a foundation, the edge types' AEBCs could prove a valuable metric in determining synthetic communication networks' wireless dependence. The high AEBCs possessed by the microwave and low-capacity radio edges could also be interpreted as an indication of potential high data transmission through the respective edge types. [9] describes high data transmission through certain edges and nodes as the leading cause of data congestion. Observing a synthesized network's edge types' AEBCs has the potential to accurately predict areas of congestion, allowing for the implementation of preventive measures at the predicted areas. The last application of AEBC would use an active network's edge types' AEBCs to assign edge types to undetermined edges in a synthesized communication network. Given the wireless edge types' high AEBC and the wired edge type's low AEBC, the edge type assignment for a selected edge in a synthetic communication network could be selected depending on the extrapolation of its AEBC magnitude.

## IV. Conclusions

This paper has presented useful statistics that could be used to generate realistic synthetic communication systems with different levels of resolution. The process to simplify the original network to get the connectivity model is also discussed. The characteristics obtained will be used to improve the model in [1], and the revised cyber network model could be emulated and studied in the testbed [10]. The detailed and simplified substation topology networks are open-sourced in [11] for further analysis.

## V. Acknowledgment

The work described in this paper was supported by funds from the US Department of Energy's DE OE0000895, the National Science Foundation's Grant 1916142, and the Sandia National Laboratories' laboratory directed research and development project #222444, titled "Harmonized Automatic Relay Mitigation of Nefarious Intentional Events (HARMONIE) Special Protection Scheme."